\documentclass[aps,prl,reprint,nofootinbib,twocolumn,showpacs,showkeys,longbibliography]{revtex4-1}

\usepackage{eurosym}
\usepackage{amsmath,amssymb,amstext}
\usepackage[usenames,dvipsnames]{color}
\usepackage{graphicx}
\usepackage{braket}
\usepackage{natbib}
\usepackage{comment}
\usepackage{dcolumn}
\usepackage[english]{babel}
\usepackage{wasysym}
\usepackage[colorlinks,bookmarks=false,citecolor=blue,linkcolor=red,urlcolor=blue]{hyperref}

\begin{document}
\title{Helical Floquet Channels in 1D Lattices} 

\author{Jan Carl Budich$^{1,2,3}$, Ying Hu$^{1,2,4,5}$, Peter Zoller$^{1,2}$}
\affiliation{$^1$Institute for Quantum Optics and Quantum Information of the Austrian Academy of Sciences, 6020 Innsbruck, Austria\\
$^2$Institute for Theoretical Physics, University of Innsbruck, 6020 Innsbruck, Austria\\
$^3$Department of Physics, University of Gothenburg, SE 412 96 Gothenburg, Sweden\\
$^4$State Key Laboratory of Quantum Optics and Quantum Optics Devices, Institute of Laser Spectroscopy, Shanxi University, Taiyuan, Shanxi 030006, China\\
$^5$Collaborative Innovation Center of Extreme Optics, Shanxi University, Taiyuan, Shanxi 030006, China}

\date{\today}
\begin{abstract}
We show how dispersionless channels exhibiting perfect spin-momentum locking can arise in a 1D lattice model. While such spectra are forbidden by fermion doubling in static 1D systems, here we demonstrate their appearance in the stroboscopic dynamics of a periodically driven system. Remarkably, this phenomenon does not rely on any adiabatic assumptions, in contrast to the well known Thouless pump and related models of adiabatic spin pumps. The proposed setup is shown to be experimentally feasible with state of the art techniques used to control ultracold alkaline earth atoms in optical lattices.   
\end{abstract}

\maketitle
\begin{figure}[h!]
\centering
\includegraphics[width=\columnwidth]{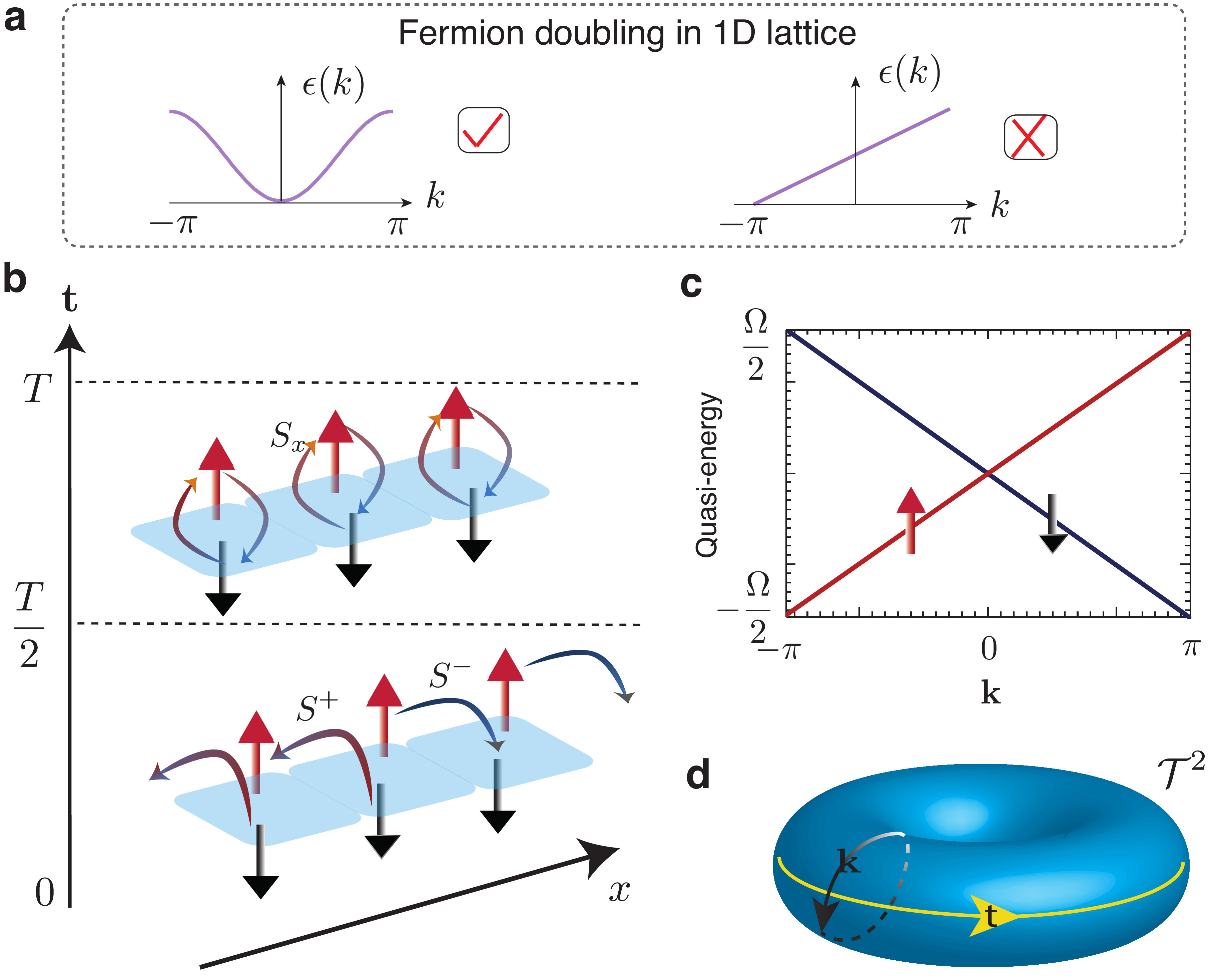}
\caption{\label{fig:one} (color online). (a) Illustration of basic constraints by fermion doubling in 1D lattice systems: The left plot shows an ordinary metallic band which must be periodic in the first Brillouin zone, while the unidirectional channel in the right plot violates this periodicity and hence is forbidden by fermion doubling. (b) Schematic of the proposed driving protocol. The spin flip hopping [see $H_1$ in Eq.~(\ref{eqn:hams})] acts during the first half-period $[0,T/2)$, while on-site spin-flips [see $H_2$ in Eq.~(\ref{eqn:hams})] characterize the second half-period $[T/2,T)$. (c) Floquet band structure of the proposed lattice model [see Eqs. (\ref{eqn:hams}-\ref{eqn:ufhelical})] with perfect spin-momentum locking. Parameters are $\alpha=\beta=\pi/T$. (d) Illustration of the toroidal time-momentum space $\mathcal T^2$.}
\end{figure}

{\textit{Introduction.}}  Exploring the rich phenomenology of spin-orbit coupling is an active field of research in numerous branches of quantum physics \cite{SpielmanReview,SOCReview,SOCPhotonics}. The discovery of helical edge-states \cite{KaneMeleQSH,Wu2006,XuMoore} has opened the route towards \textit{perfect spin-momentum locking}, characterized by a one-to-one correspondence between the propagation direction of particles and their spin. Such exotic states have only been realized at the surface of 2D topological insulators \cite{KaneMeleQSH,BHZ,QSHExperiment,HasanKane2010,QiReview2011}. Without the 2D bulk, their occurrence is forbidden in 1D lattice systems \cite{QiReview2011}, as the periodicity of band structures in the first Brillouin Zone (BZ) imposes fundamental constraints -- referred to as fermion doubling \cite{FermionDoubling} [cf.~Fig.~1(a)]. Harnessing the unique properties of periodically driven quantum systems \cite{Rudner2013,Goldman2014,Klinovaja2016,Lindner2016,Moessner2016,Vishwanath2016}, here we show how these limitations can be circumvented: we find perfect spin-momentum locking in the \textit{stroboscopic} dynamics of a periodically driven 1D lattice model. While conventional helical edge states require a time reversal symmetric topological 2D bulk \cite{KaneMeleZ2}, the spin-momentum locking in our 1D setting stems from topological properties in combined time-momentum (Floquet) space [see Fig.~\ref{fig:one}(d)], and relies on a spin-rotation symmetry of the stroboscopic dynamics. Our approach goes conceptually beyond adiabatically projected models such as the Thouless pump \cite{ThoulessPump,BlochThoulessPump}, in that we consider the full quasi-energy spectrum \textit{without} involving adiabatic projections.

In Floquet systems, the quasi-energies are only defined modulo the driving frequency $\Omega$, allowing for spectra that are only periodic in the BZ up to integer multiples of $\Omega$. However, even in driven systems, unidirectional motion in 1D systems cannot be achieved {\textit{without adiabatic assumptions}}, due to fundamental topological constraints \cite{Kitagawa2010}. The central result of this work is that the Floquet Bloch Hamiltonian ($\hbar=1$)
\begin{align}
H^F=v\, k\, \sigma_z
\label{eqn:ksz}
\end{align}
exhibiting the perfect spin-momentum locking [see Fig.~\ref{fig:one}(c)] familiar from the helical edge states of 2D topological insulators can still be achieved in a {\textit{microscopic 1D lattice}} model. Eq.~(\ref{eqn:ksz}), with the lattice momentum $k$ and standard Pauli matrix $\sigma_z$, describes two spin-species that independently perform an opposite {\textit{uni-directional}} and {\textit{dispersionless}} motion with velocity $v$.

\begin{figure}[h!]
\centering
\includegraphics[width=\columnwidth]{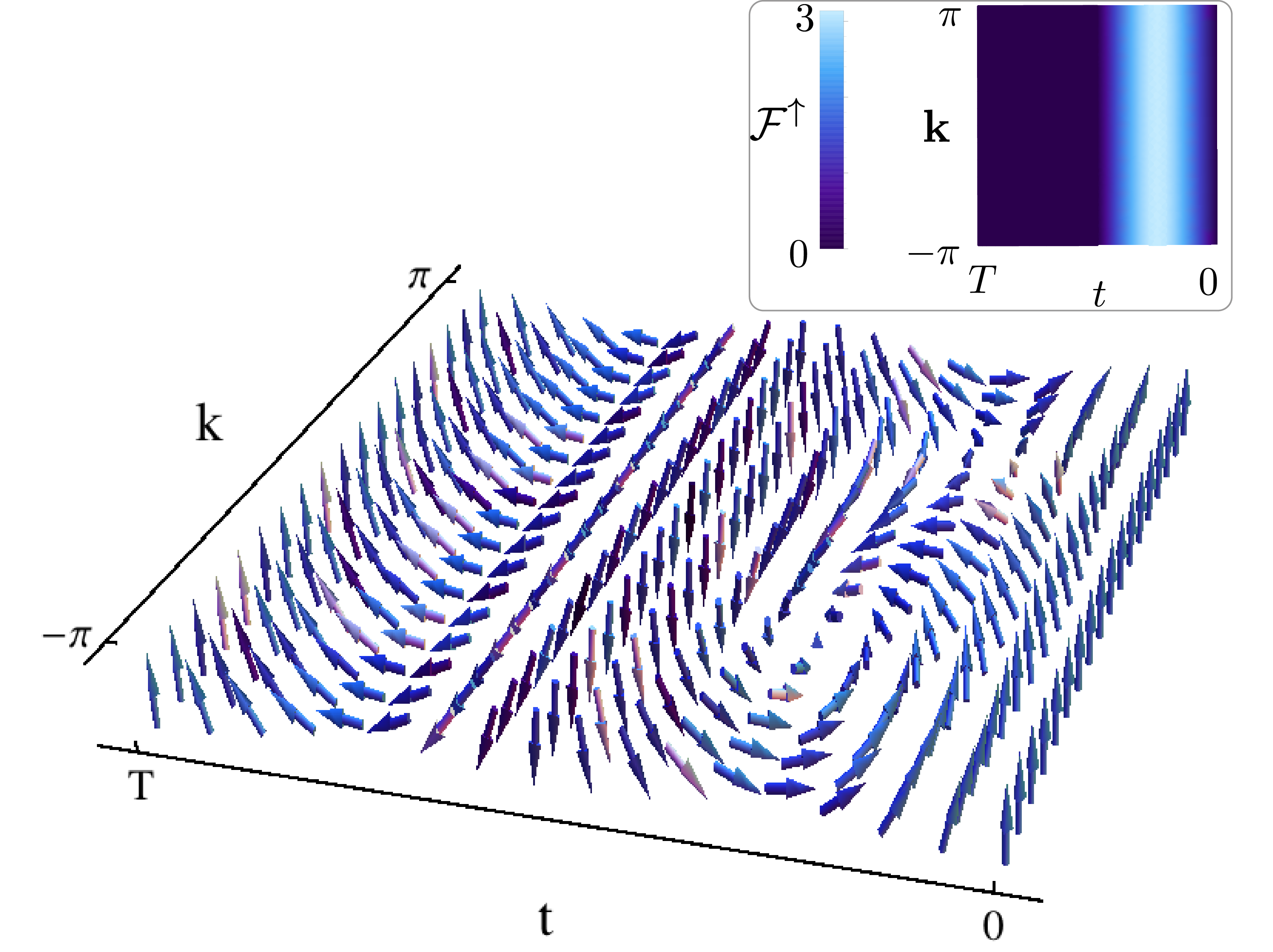}
\caption{\label{fig:two} (color online). Lower panel: Topologically non-trivial spin micro-motion of the Bloch states $\lvert u_k^\uparrow(t)\rangle=U_k(t,0)\lvert u_k^\uparrow(0)\rangle$ that are eigenstates of the spin up Floquet operator $U_k^\uparrow(T,0)$ at stroboscopic times $t=0~(\text{mod}~T)$. Upper panel: Berry curvature $\mathcal F^\uparrow_{k,t}=2\text{Im}\left\{\langle \partial_ku_k^\uparrow(t)\vert \partial_t u_k^\uparrow(t)\rangle\right\}$ in combined time-momentum space. Parameters are $\alpha=\beta=\pi/T=\pi$ in all plots.}
\end{figure}

Remarkably, such a behavior is possible beyond adiabatic dynamics even though the uni-directional motion of a single spin-species cannot be achieved. To see this, we note that $H^F$ in a Floquet system generates the stroboscopic dynamics described by the time evolution operator $U(T,0)=\text{e}^{-iH^FT}$ over one driving period $[0,T)$ with $T=\frac{2\pi}{\Omega}$. During the so-called micro-motion within a period, the two spin-species are necessarily intertwined in a topologically non-trivial fashion [see Fig.~\ref{fig:one}(b) and Fig.~\ref{fig:two}] as we discuss below. In addition, we provide an experimentally feasible proposal for realizing this scenario with ultracold alkaline earth atoms (AEAs) in optical lattices.
\\

{\textit{Lattice model with perfect spin-momentum locking. }}
We consider a Floquet system of fermions with spin $1/2$ annihilated by the spinor operators $\psi_j=(\psi_{j\uparrow},\psi_{j\downarrow})$ on a 1D lattice with unit lattice constant. The driving protocol consists of switching between two non-commuting time-independent Hamiltonians $H_1$ and $H_2$, such that $H_1$ generates the time-evolution during the first half-period $[0,T/2)~(\text{mod}~ T)$ whereas $H_2$ operates during the second half-period  $[T/2,T)~(\text{mod}~ T)$. The explicit form of $H_1$ and $H_2$ reads as [see Fig.~\ref{fig:one}(b) for an illustration]
\begin{align}
H_1=-\alpha \! \sum_j\psi_j^\dag S^+\psi_{j+1}+\text{h.c.},\quad \!\!\!H_2=\beta\!\sum_j\psi_j^\dag \sigma_x \psi_j,
\label{eqn:hams}
\end{align}
where $S^+=\frac{1}{2}(\sigma_x+i\sigma_y)$ flips the spin from down to up and $\alpha,\beta$ are real coupling constants \cite{footbeta}. Both $H_1$ and $H_2$ are lattice translation invariant rendering the lattice momentum $k$ a good quantum number and allowing us to factorize the time-evolution operator into momentum components $U_k(t,t_0)$. For the parameter choice $\alpha=\beta=\pi/T$, we obtain
\begin{align}
U_k(T,0)=\text{e}^{-iH_2^kT/2}\text{e}^{-iH_1^kT/2}=\text{e}^{-ik\sigma_z},
\label{eqn:ufhelical}
\end{align}
where $H_1^k=-\alpha [\cos(k)\sigma_x-\sin(k)\sigma_y]$ and $H_2^k=\beta \sigma_x$ are the Bloch Hamiltonians associated with $H_1$ and $H_2$ [see Eq.~(\ref{eqn:hams})], respectively. Computing the associated Floquet Bloch Hamiltonian $H_k^F=(i/T)\log[U_k(T,0)]=(1/T)k\sigma_z$, we recover Eq.~(\ref{eqn:ksz}) with the velocity $v=1/T$. Note that $H_k^F$, when interpreted as a static Bloch Hamiltonian, contains a discontinuous jump at $k=\pm \pi$ and hence cannot be achieved by any (local) hopping in real space. Quite remarkably, in the present Floquet setting, it can be achieved - or practically at least be arbitrarily closely approached -  by simply tuning the parameters $\alpha$ and $\beta$ in the local instantaneous Hamiltonians (\ref{eqn:hams}).

The spin-momentum locking in the proposed Floquet system may be understood at an intuitive level [see also Fig.~\ref{fig:one}(b)]. The Hamiltonian $H_1$ drives a hopping process between nearest neighbor sites during the first half-period, where opposite directions of motion are tied to opposite spin-flip operations. However, once a particle has hopped, it has the wrong spin to hop into the same direction again, since $(S^+)^2=0$. To prevent this deadlock, $H_2$ recharges the spin-pump during the second half-period by flipping back the spin on-site. Putting together the two half-cycles, each particle has moved by one site with a perfect spin-momentum locking after a full period.\\

 {\textit{Implementation with alkaline earth atoms. }}
The lattice model (\ref{eqn:hams}-\ref{eqn:ufhelical}) may be experimentally implemented with state of the art techniques for the control of ultracold atoms [see Ref.~\cite{BlochReview} for a review], where Raman processes are employed to design laser-assisted hopping in optical lattices \cite{JakschZollerNJP,GerbierDalibard,DalibardReview,Marie2015,Spielman2015}.  An ideal experimental platform in this context is provided by gases of AEAs such as Yb [see, e.g. Ref.~\cite{Marie2015}]. There, the spin degree of freedom $\sigma$ occurring in our model is encoded in two Zeeman levels with different magnetic quantum number $m_F$ of the atomic ground state of $^{173}$Yb. Spin-flip processes are then controlled by optical dipole selection rules of the involved Raman transitions.  A detailed proposal for the implementation of the spin-flip hopping characterizing $H_1$ based on the experimental tools of Ref.~\cite{Marie2015} has recently been published \cite{CatherineHelical}. The on-site spin flip processes defining $H_2$ have already been extensively employed experimentally \cite{Marie2015} to realize hopping in so called synthetic dimensions \cite{Celi2014,GoldmanReview}, where internal states of the atom are interpreted as lattice sites in an extra dimension. To experimentally realize our two-step driving protocol [see Fig.~\ref{fig:one}(b)], we propose to use pulsed Raman lasers switching between laser-assisted spin-flip hopping ($H_1$) and on-site spin-flips ($H_2$). An alternative implementation of our model may be provided by a superlattice setting with double-well super-sites encoding the spin degrees of freedom, which can be readily implemented using alkali atoms \cite{BlochThoulessPump}.

\begin{figure}[h!]
\centering
\includegraphics[width=\columnwidth]{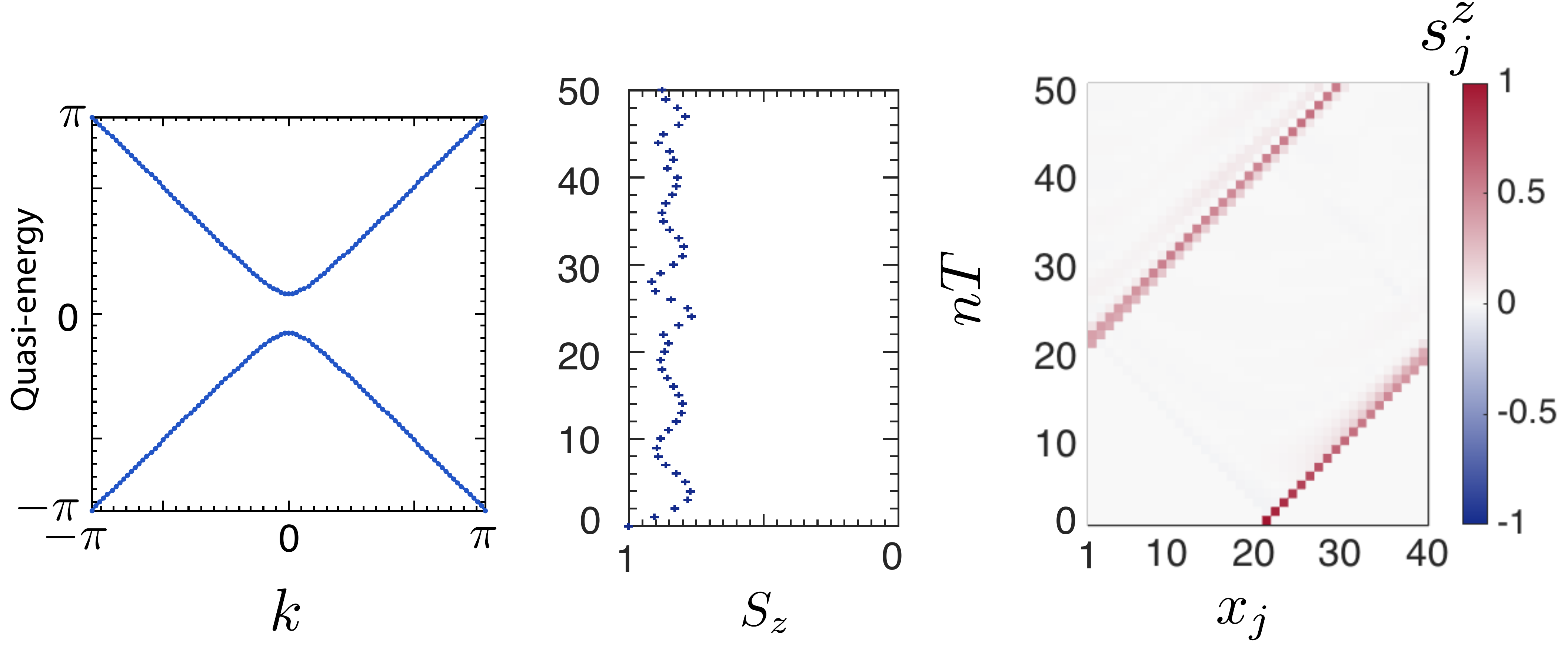}
\includegraphics[width=\columnwidth]{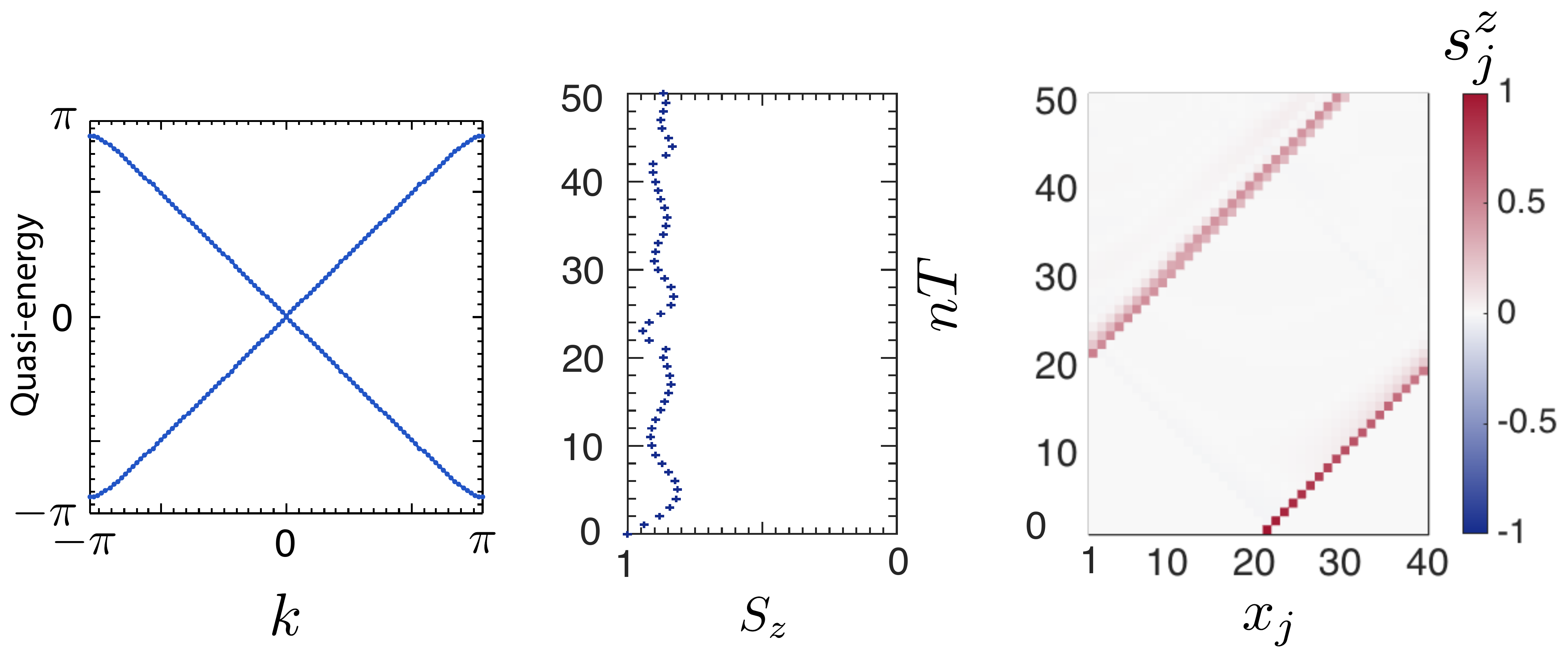}
\caption{\label{fig:perfect} (color online). Top: Gap around $k=0$ for $\alpha=1.1\pi/T,\beta=0.9\pi/T$. From left to right, the three plots show the Floquet spectrum of the system, the total $S_z$ polarization as a function of time, and the spatially resolved $S_z$ polarization as a function of time. Bottom: Gap around $k=\pi$ for $\alpha=\beta=0.92 \pi/T$. The plots are analogous to those in the top panel.}
\end{figure}

{\textit{Topological analysis. }} We now provide a deeper understanding in terms of topology of how the phenomenology discussed above can arise in a microscopic lattice model without relying on adiabatic projections. We stress the different role of topology in our present setting, as compared to conventional helical edge states. In 2D topological insulators, a topological invariant associated with the time-reversal invariant insulating bulk of the system entails and protects the presence of helical edge states \cite{KaneMeleQSH,QiReview2011}. Here, instead an emergent spin-rotation symmetry in the stroboscopic dynamics of the 1D system allows for the definition of a topological invariant that entails and protects helical Floquet modes as described by Eq.~(\ref{eqn:ksz}). The protecting symmetry of the Floquet spectrum (Floquet symmetry) in our model (\ref{eqn:hams}) requires tuning the system to the parameter line $\alpha=\beta=\pi/T$. However, below we show with numerical simulations [see Fig.~\ref{fig:perfect}] that, even in the presence of significant deviations from this ideal situation, clear signatures of the spin-momentum locking are still experimentally observable. The Floquet operator $U_k(T,0)$ in Eq.~(\ref{eqn:ufhelical}) with $\alpha=\beta=\pi/T$ preserves $S_z=\sigma_z/2$ and can hence be decomposed into two irreducible blocks $U_k^\sigma(T,0),~\sigma=\uparrow,\downarrow$. The Floquet winding number \cite{Kitagawa2010} for the individual spin blocks reads as
\begin{align}
\nu_\sigma=\frac{1}{2\pi i}\oint_{\text{BZ}}\text{d}k\,\text{Tr}[U_k^\sigma\partial_k {U_k^\sigma}^\dag]=\frac{1}{\Omega}\oint_{\text{BZ}}\text{d}k\,\sum_\alpha(\partial_k \epsilon_k^{\sigma,\alpha})
\label{eqn:nusigma}
\end{align}
with the Floquet quasi-energies $\epsilon_k^{\sigma,\alpha}$ for band $\alpha$ in spin block $\sigma$. We note that in our specific model, there is only one band per spin block. The topological invariant $\nu_\sigma$ simply counts the number of chiral Floquet modes with spin $\sigma$, i.e. Floquet bands which are periodic in the BZ only modulo $\Omega$. For the model in Eq.~(\ref{eqn:ufhelical}), $\nu_\sigma=\pm 1$ for $\sigma=\uparrow,\downarrow$. 

In Ref.~\cite{Kitagawa2010}, a similar Floquet winding number $\nu$ has been introduced, counting the total number of chiral Floquet modes without assuming a spin-rotation symmetry. Furthermore, it has been shown that $\nu$ is identical to the Chern number \cite{Chern,Thouless1982} of the 2D system characterized by the Bloch functions $\lvert u_k^\alpha(t)\rangle=U_k(t,0)\lvert u_k^\alpha(0)\rangle$ in combined $(k,t)$ space [see Fig.\ref{fig:one}(d)], where $\alpha$ labels the Floquet Bloch bands and $\lvert u_k^\alpha(0)\rangle$ is family of eigenfunctions of the Floquet operator $U_k(T,0)$. This relation implies that a non-zero $\nu$ can only occur in effective models such as the Thouless pump \cite{ThoulessPump}, where some energetically higher-lying bands have been adiabatically eliminated before computing the Floquet quasi band structure. This is because the Chern numbers of all bands obey a \textit{zero sum rule} in lattice models. The intuitive picture behind this rule is that the Chern number of a subspace with projection $P(k,t)=\sum_\alpha \lvert u_k^\alpha(t)\rangle\langle u_k^\alpha(t)\rvert$ measures the winding of the orientation of this subspace in the total Hilbert space. If the considered Floquet system contains all bands, we have $P(k,t)=\mathbf 1$ and no non-trivial winding is possible.

By contrast, in our microscopic lattice model (\ref{eqn:hams}-\ref{eqn:ufhelical}), a non-trivial $\nu_\sigma$ is possible because the two spin species are intertwined during the micro-motion, i.e., by the time-evolution operator $U_k(t,0),~0<t<T$. The resulting winding in spin space of the Bloch functions $\lvert u_k^\uparrow(t)\rangle=U_k(t,0)\lvert u_k^\uparrow(0)\rangle$ with $\lvert u_k^\uparrow(0)\rangle$ denoting an eigenfunction of $U_k^\uparrow(T,0)$ is shown in the lower panel of Fig.~\ref{fig:two}. The Berry curvature $\mathcal F^\uparrow_{k,t}=2\text{Im}\left\{\langle \partial_ku_k^\uparrow(t)\vert \partial_t u_k^\uparrow(t)\rangle\right\}$ is shown in the upper panel of Fig.~\ref{fig:two}. Computing the Chern number $\mathcal C^\uparrow$ associated with the toroidal combined momentum-time space $\mathcal T^2$ [see Fig.~\ref{fig:one} (d)] yields $\mathcal C^\uparrow=(1/2\pi)\int_{\mathcal T^2}\mathcal F^\uparrow=\nu_\uparrow=1$, and, in agreement with the mentioned zero sum rule of Chern numbers $\mathcal C^\downarrow=\nu_\downarrow=-1$.\\

{\textit{Stability of spin-momentum locking. }} 
{We now show that the spin momentum locking stays robust and clearly observable even in the presence of deviations from the parameter line $\alpha=\beta=\pi/T$ representing possible experimental imperfections}. 

We first study the visibility of the spin-momentum locking for a localized wave-packet initialized at site $j=21$ with spin up polarization. We numerically simulate a system with a size of $L=40$ lattice sites. In the following, we focus on periodic boundary conditions, noting that open boundary conditions simply lead to a perfect reflection of the particles involving a spin-flip on the outermost sites. In Fig.~\ref{fig:perfect}, we summarize our results if (i) a gap around $k=0$ is opened in the quasi-energy spectrum by setting $\alpha\ne\beta$ [see top panel], and (ii) if a gap is opened around $k=\pm\pi$ for $\alpha=\beta\ne \pi/T$ [see bottom panel]. The effects of such imperfections are twofold.  First, due to the deviation from a perfectly linear dispersion, the initially sharply localized wave-packet slightly spreads out in real space. Second, due to a coupling of the two spins, a finite spectral weight of the opposite spin species ($<5\%$ for a relative deviation of $10\%$ in the system parameters) is generated. Our numerical data shows that the spin-momentum locking is still clearly visible, even for significant deviations from the ideal parameter line $\alpha=\beta=\pi/T$.   

Generally speaking, in the presence of symmetry breaking imperfections, a gap may open around $\Omega/2$ in the quasi-energy spectrum. However, when interpreting the resulting $H_k^F$ as a static band structure, it would still be extremely challenging to realize, as the corresponding decay length of the hopping range in real space {\emph{diverges}} on approaching the parameter line $\alpha=\beta=\pi/T$. Instead, in the present Floquet scheme, an arbitrarily non-local $H_k^F$ exhibiting arbitrarily precise spin momentum locking can readily be experimentally achieved by (approximately) tuning the local coupling strengths $\alpha$ and $\beta$.

\begin{figure}[h!]
\centering
\includegraphics[width=\columnwidth]{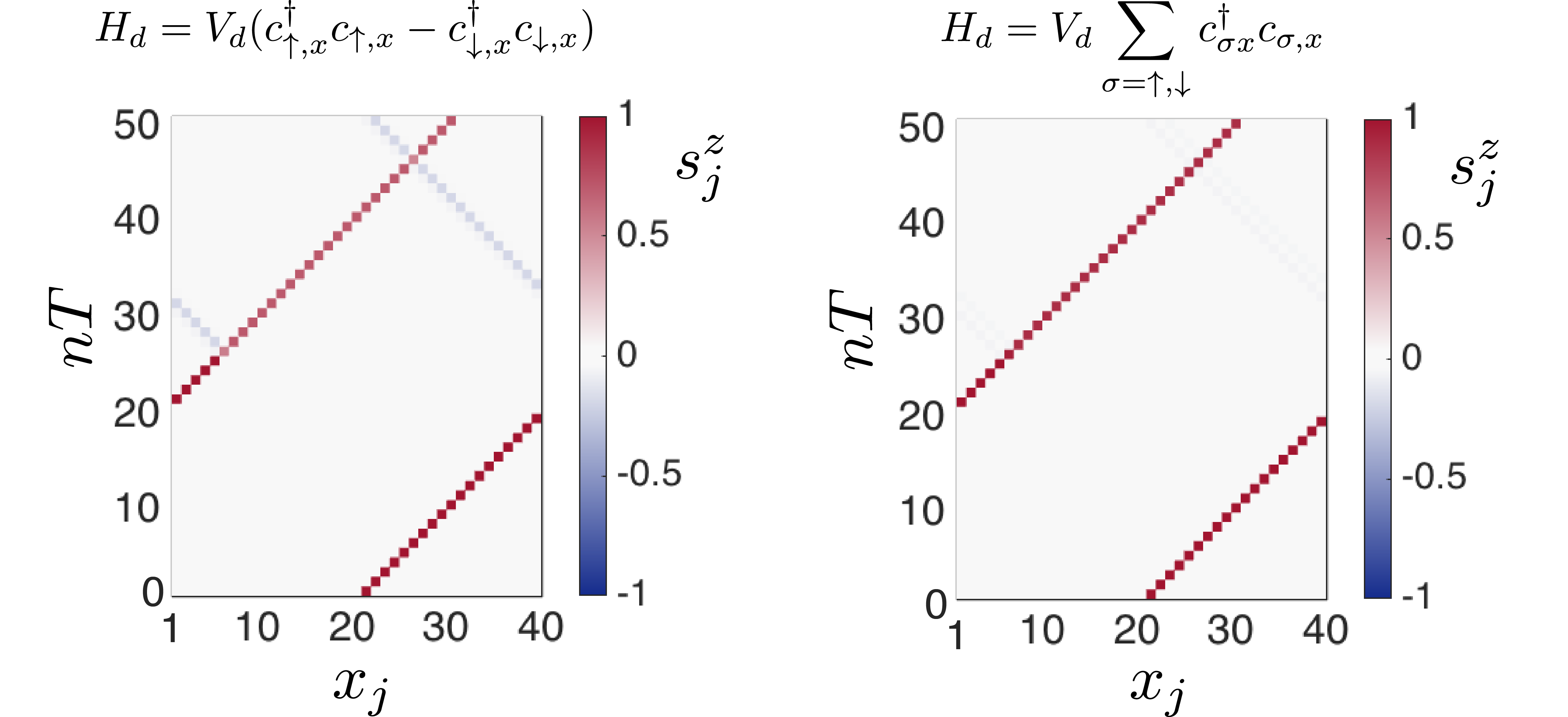}
\caption{\label{fig:impurity} (color online). Left: Scattering due to $\sigma_z$ impurity a site $x=6$. The plot shows the spatial distribution of the $S_z$ polarization as a function of time. Right: Scattering due to $\sigma_0$ impurity at site $x=6$. The bulk parameters are $\alpha=\beta=\pi/T$ and $V_d=1.5/T$ in both plots.}
\end{figure}

In addition, we study the influence of various imperfections that break the translation-invariance in our system [see Fig.~\ref{fig:impurity}]. Specifically, we consider a single spin-dependent impurity of strength $V_d$ at site $x$ modelled by the Hamiltonian $H_d=V_d(c_{\uparrow,x}^\dag c_{\uparrow,x}-c_{\downarrow,x}^\dag c_{\downarrow,x})$ [see Fig.~\ref{fig:impurity} left panel], and a spin-independent impurity modelled by the Hamiltonian  $H_d=V_d\sum_\sigma c_{\sigma,x}^\dag c_{\sigma,x}$ [see Fig.~\ref{fig:impurity} right panel]. The spin-independent impurity does not have a strong influence on the dynamics of the wave packet, even for an impurity strength $V_d=1.5/T$. By contrast, the spin-dependent impurity is found to cause significant scattering, but the scattered wave packet has both reversed direction of motion and reversed spin-polarization, thus keeping the spin-momentum locking intact.\\ 

{\textit{Concluding discussion. }} 
For periodically driven 2D systems, it has recently been shown \cite{Rudner2013} how chiral edge states can occur, even if all quasi-energy bands are characterized by a zero Chern number -- a no go for static systems. In our present work, even without any 2D bulk, we have found a 1D Floquet counterpart [see Eq.~(\ref{eqn:ksz})] of helical edge states known from 2D topological insulators. Since Eq.~(\ref{eqn:ksz}) cannot be realized as a local Hamiltonian in a \textit{static} microscopic 1D lattice model, our results give a new intriguing example of how periodically driven systems can dynamically enable the realization of exotic states of matter. Remarkably, the microscopic model (\ref{eqn:hams}) and driving protocol proposed here is of immediate experimental relevance as it can be implemented by combining state of the art techniques to trap and manipulate ultracold quantum gases. 

We note that a unidirectional motion has been recently realized \cite{WhiteQuantumWalk,Kitagawa2012} in quantum walk setups \cite{KitagawaQuantumWalk,Tarasinski2014,Asboth2015} in a photonic context. There, the essential physical mechanism relies on the higher spatial dimension of the setup: A beam displacer \textit{redirects} the uni-directional motion of  the incident laser beam into a step of the walk in a perpendicular direction. By contrast, here we are interested in a fermionic quantum many-body system in a microscopic 1D lattice potential, where the dynamics is constrained by fermion doubling. In an atomic setup, a unidirectional quantum walk has been engineered based on the adiabatic modulation of spin-dependent lattice potentials (see, e.g., \cite{Karski2009,Craig2014}),  while our present driving protocol is based on a stationary lattice potential and does not rely on adiabatic assumptions.\\

In a broader context, helical channels have been identified as promising candidates for numerous applications. In the field of spintronics, their perfect spin-momentum locking may enable new possibilities to control spin properties by all electric means. Regarding the realization of exotic quasi-particles, hybrid systems involving helical channels coupled to superconductors have repeatedly appeared, both in the context of Majorana bound states \cite{FuKane2008} and, more recently, in the theoretical prediction of fractional Majorana fermions or parafermions in strongly correlated systems \cite{ZhangKane, Orth2015, AliceaFendley}. The Floquet counterpart of helical channels reported in our present work may be of key interest along these lines: First, from a computational perspective, our microscopic 1D lattice model  model (\ref{eqn:hams}) will even in the presence of pairing terms and correlations still be amenable to first principle numerical analysis, e.g. by means of time-dependent density matrix renormalization group techniques. Second, the inherently time-dependent character of the proposed system may lead to phenomena in such hybrid systems [see e.g. Ref.~\cite{FloquetMajorana} for the example of Floquet Majorana states at finite quasi-energy] that are not found in their static counterparts. Finally, the simplicity and  feasibility of our proposal hold great promise for the observation of such new physics in future experiments.  

{\textit{Acknowledgment.}} This project was supported by the ERC Synergy Grant UQUAM and the SFB FoQuS (FWF Project No. F4016-N23). Y. H. also acknowledges the support from the Institut f\"{u}r Quanteninformation GmbH.

\bibliographystyle{apsrev}

\begin{thebibliography}{14}
\bibitem{footbeta}
We note that a complex $\alpha$ simply amounts to a global shift in the lattice momentum.
\end{thebibliography}

\end{document}